\documentclass[11pt,twoside]{article}

\usepackage{graphicx}
\usepackage{amssymb}
\usepackage{asp2006}
\usepackage{epsf}
\usepackage{lscape}

\newcommand{\mbh}{M_{\bullet}}
\newcommand{\msun}{M_{\sun}}

\newcommand{\mcnd}{M_\mathrm{CND}}
\newcommand{\rcnd}{R_\mathrm{CND}}

\newcommand{\pc}{\mathrm{pc}}
\newcommand{\myr}{\mathrm{Myr}}

\newcommand{\rd}{\mathrm{d}}

\begin{document}
\title{Stellar disc -- dynamical evolution in a perturbed potential}
\author{Ladislav \v{S}ubr}   
\affil{Astronomical Institute, Charles University in Prague}

\begin{abstract} 
Models of the origin of young stars in the Galactic
Centre are facing various problems. The most promissing scenario of the
star formation in a thin self-gravitating disc naturally forms stars on
coherently rotating orbits, but it fails to explain
origin of several tens of stars that evidently do not belong to any of
the disc-like structures in the GC. One possible solution lies in rather
complicated initial conditions, assuming at least two infalling and interacting
gas clouds. We present alternative solution showing that a single thin stellar
disc may have given birth to all young stars in the GC. The outliers are
explained as stars that have been stripped from the parent structure
due to the gravitational interaction with the gaseous circum-nuclear disc.
\end{abstract}


We investigate a model of the Galactic Centre which contains following
constituents:
{\bf{}Supermassive black hole (SMBH)} of mass $\mbh = 3.5\times10^6\msun$
dominates the gra\-vi\-ta\-tional field and it is approximated by a fixed
Keplerian potential.
{\bf{}Circum-nuclear Disc (CND)} is modelled by several tens of particles
orbiting the SMBH in a toroidal structure. Its total mass is not well
determined by observations; we assume $\mcnd \approx 0.2\mbh$
orbiting at a characteristic radius of $\rcnd \approx 1.5\pc$.
{\bf{}Stellar cusp} of mass $\gtrsim0.1\mbh$ within $1\pc$ is assumed to be
spherically symmetric. We incorporate it by means of a smooth fixed potential.
{\bf{}Young stellar disc} is a set of $\sim100$ equal-mass gravitating particles.
Initially, they are on circular orbits with nearly colinear angular momenta.
Semi-major axes have distribution $\propto a^{-1}$
within the interval $\langle 0.04\pc, 0.4\pc \rangle$.

\begin{figure}
\includegraphics[width=\textwidth]{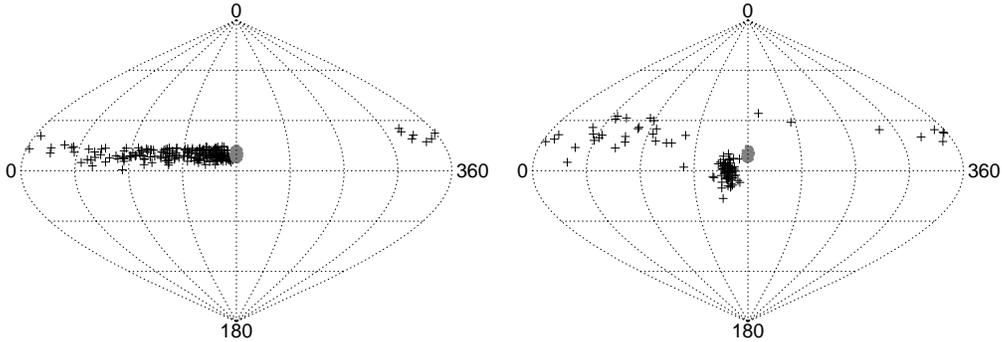}
\caption{Dissolution of the coherently rotating structure due to the gravity
of stellar cusp and the CND in terms of the direction of the angular momenta.
Self-gravity of the stellar disc is considered
only in the right panel, while it is omitted in
the left one. Normal vectors of all orbits lie within the shaded area
at $T=0$; crosses represent their position at $T=7\myr$. Coordinate system
is aligned with the axis of the CND.}
\label{fig:dissolve}
\end{figure}

Provided the CND would be the only perturbation to the SMBH's potential,
individual stellar orbits would undergo secular evolution.
In particular, orbital eccentricity and inclination would oscillate on the
time-scale comparable to the age of the young stars.
These oscillations are considerably damped if additional {\em spherically\/}
symmetric perturbation of the late type stellar cusp is present. This
component, however, does not suppress
differential precession of the stellar orbits around the symmetry axis of
the CND. The precession rate can be approximated as \cite{subr09}
$$
 \frac{\rd\Omega}{\rd t} \approx -{\textstyle\frac{3}{4}}\,
 \cos i \, \frac{\mcnd}{\mbh} \, \frac{a\sqrt{G\mbh a}}{\rcnd^3} \,
 \frac{1+\frac{3}{2}e^2}{\sqrt{1-e^2}} \approx const\,.
$$
There is a strong dependence of $\rd\Omega / \rd t$ on the semi-major
axis and the orbit inclination $i$ with respect to the plane of the CND.
Stars at the outer part of the disc and/or stars that are less perpendicular
to the CND undergo faster precession, i.e. the disc becomes warped. The shape
of the stellar disc in terms of the normal
vectors of the orbits after several Myrs of such a differential precession
is shown in the right panel of Fig.~\ref{fig:dissolve}.
We have verified \cite{subr09} that such a model of a warped stellar disc is
compatible with the observational data available in the literature
\cite{paumard06,bartko09}.
The picture changes when self-gravity of the stellar disc is taken into
account (see Fig.~\ref{fig:dissolve}). It appears that mutual torques of the
stellar orbits tend to drive
the coherently rotating `core' of the disc towards the orientation perpendicular
to the CND. At the same time, two-body relaxation scatters some stars to orbits
that undergo fast precession and do not apparently belong to the parent structure
after few Myrs.
Another consequence of the gravitational interaction
of the stellar orbits with both the CND and stellar disc itself is an
enhanced growth of orbital eccentricities \cite{haas10}.

\section*{Conclusions}
We suggest that all young stars (probably except for S-stars) in the Galactic
Centre could have been formed in a single thin self-gravitating disc.
Perturbative influence of the CND then led to partial destruction of the
coherently rotating structure. In addition to our previous results, we have
found that the core of the disc has tendency to migrate towards orientation
perpendicular to the CND, which is in a striking accord with the data provided
by the observations. This result cannot be achieved if the gravity of either
the CND, the spherical cusp or the stellar disc is ignored.

\acknowledgements 
This work was supported by the Czech Science Foundation (ref.\ 205/07/0052)
and the Centre for Theoretical Astrophysics in Prague.


\end{document}